\documentstyle[twoside,psfig]{article}

\catcode`\@=11
\long\def\@makefntext#1{
\protect\noindent \hbox to 3.2pt {\hskip-.9pt  
$^{{\eightrm\@thefnmark}}$\hfil}#1\hfill}		

\def\@makefnmark{\hbox to 0pt{$^{\@thefnmark}$\hss}}	
	
\def\ps@myheadings{\let\@mkboth\@gobbletwo
\def\@oddhead{\hbox{}
\rightmark\hfil\eightrm\thepage}   
\def\@oddfoot{}\def\@evenhead{\eightrm\thepage\hfil
\leftmark\hbox{}}\def\@evenfoot{}
\def\sectionmark##1{}\def\subsectionmark##1{}}

\oddsidemargin=\evensidemargin
\newcounter{sectionc}\newcounter{subsectionc}\newcounter{subsubsectionc}
\renewcommand{\section}[1] {\vspace{12pt}\addtocounter{sectionc}{1} 
\setcounter{subsectionc}{0}\setcounter{subsubsectionc}{0}\noindent 
	{\tenbf\thesectionc. #1}\par\vspace{5pt}}
\renewcommand{\subsection}[1] {\vspace{12pt}\addtocounter{subsectionc}{1} 
	\setcounter{subsubsectionc}{0}\noindent 
	{\bf\thesectionc.\thesubsectionc. {\kern1pt \bfit #1}}\par\vspace{5pt}}
\renewcommand{\subsubsection}[1] {\vspace{12pt}\addtocounter{subsubsectionc}{1}
	\noindent{\tenrm\thesectionc.\thesubsectionc.\thesubsubsectionc.
	{\kern1pt \tenit #1}}\par\vspace{5pt}}
\newcommand{\nonumsection}[1] {\vspace{12pt}\noindent{\tenbf #1}
	\par\vspace{5pt}}

\newcounter{appendixc}
\newcounter{subappendixc}[appendixc]
\newcounter{subsubappendixc}[subappendixc]
\renewcommand{\thesubappendixc}{\Alph{appendixc}.\arabic{subappendixc}}
\renewcommand{\thesubsubappendixc}
	{\Alph{appendixc}.\arabic{subappendixc}.\arabic{subsubappendixc}}

\renewcommand{\appendix}[1] {\vspace{12pt}
        \refstepcounter{appendixc}
        \setcounter{figure}{0}
        \setcounter{table}{0}
        \setcounter{lemma}{0}
        \setcounter{theorem}{0}
        \setcounter{corollary}{0}
        \setcounter{definition}{0}
        \setcounter{equation}{0}
        \renewcommand{\thefigure}{\Alph{appendixc}.\arabic{figure}}
        \renewcommand{\thetable}{\Alph{appendixc}.\arabic{table}}
        \renewcommand{\theappendixc}{\Alph{appendixc}}
        \renewcommand{\thelemma}{\Alph{appendixc}.\arabic{lemma}}
        \renewcommand{\thetheorem}{\Alph{appendixc}.\arabic{theorem}}
        \renewcommand{\thedefinition}{\Alph{appendixc}.\arabic{definition}}
        \renewcommand{\thecorollary}{\Alph{appendixc}.\arabic{corollary}}
        \renewcommand{\theequation}{\Alph{appendixc}.\arabic{equation}}
        \noindent{\tenbf Appendix \theappendixc #1}\par\vspace{5pt}}
\newcommand{\subappendix}[1] {\vspace{12pt}
        \refstepcounter{subappendixc}
        \noindent{\bf Appendix \thesubappendixc. {\kern1pt \bfit #1}}
	\par\vspace{5pt}}
\newcommand{\subsubappendix}[1] {\vspace{12pt}
        \refstepcounter{subsubappendixc}
        \noindent{\rm Appendix \thesubsubappendixc. {\kern1pt \tenit #1}}
	\par\vspace{5pt}}

\newcommand{\textlineskip}{\baselineskip=13pt}
\newcommand{\smalllineskip}{\baselineskip=10pt}

\def\eightcirc{
\begin{picture}(0,0)
\put(4.4,1.8){\circle{6.5}}
\end{picture}}
\def\eightcopyright{\eightcirc\kern2.7pt\hbox{\eightrm c}} 

\newcommand{\copyrightheading}[1]
	{\vspace*{-2.5cm}\smalllineskip{\flushleft
	{\footnotesize Modern Physics Letters A, #1}\\
	{\footnotesize $\eightcopyright$\, World Scientific Publishing
	 Company}\\
	 }}

\newcommand{\publisher}[2]{{\begin{center}\footnotesize\smalllineskip 
	Received #1\\
	Revised #2
	\end{center}
	}}

\def\abstracts#1#2#3{{
	\centering{\begin{minipage}{4.5in}\footnotesize\baselineskip=10pt
	\parindent=0pt #1\par 
	\parindent=15pt #2\par
	\parindent=15pt #3
	\end{minipage}}\par}} 

\def\keywords#1{{
	\centering{\begin{minipage}{4.5in}\footnotesize\baselineskip=10pt
	{\footnotesize\it Keywords}\/: #1
	 \end{minipage}}\par}}

\newcommand{\bibit}{\nineit}
\newcommand{\bibbf}{\ninebf}
\renewenvironment{thebibliography}[1]
	{\frenchspacing
	 \ninerm\baselineskip=11pt
	 \begin{list}{\arabic{enumi}.}
        {\usecounter{enumi}\setlength{\parsep}{0pt}     
	 \setlength{\leftmargin 12.7pt}{\rightmargin 0pt} 
         \setlength{\itemsep}{0pt} \settowidth
	{\labelwidth}{#1.}\sloppy}}{\end{list}}

\newcounter{itemlistc}
\newcounter{romanlistc}
\newcounter{alphlistc}
\newcounter{arabiclistc}

\newcommand{\fcaption}[1]{
        \refstepcounter{figure}
        \setbox\@tempboxa = \hbox{\footnotesize Fig.~\thefigure. #1}
        \ifdim \wd\@tempboxa > 5in
           {\begin{center}
        \parbox{5in}{\footnotesize\smalllineskip Fig.~\thefigure. #1}
            \end{center}}
        \else
             {\begin{center}
             {\footnotesize Fig.~\thefigure. #1}
              \end{center}}
        \fi}

\newcommand{\tcaption}[1]{
        \refstepcounter{table}
        \setbox\@tempboxa = \hbox{\footnotesize Table~\thetable. #1}
        \ifdim \wd\@tempboxa > 5in
           {\begin{center}
        \parbox{5in}{\footnotesize\smalllineskip Table~\thetable. #1}
            \end{center}}
        \else
             {\begin{center}
             {\footnotesize Table~\thetable. #1}
              \end{center}}
        \fi}

\def\@citex[#1]#2{\if@filesw\immediate\write\@auxout
	{\string\citation{#2}}\fi
\def\@citea{}\@cite{\@for\@citeb:=#2\do
	{\@citea\def\@citea{,}\@ifundefined
	{b@\@citeb}{{\bf ?}\@warning
	{Citation `\@citeb' on page \thepage \space undefined}}
	{\csname b@\@citeb\endcsname}}}{#1}}

\newif\if@cghi
\def\cite{\@cghitrue\@ifnextchar [{\@tempswatrue
	\@citex}{\@tempswafalse\@citex[]}}
\def\citelow{\@cghifalse\@ifnextchar [{\@tempswatrue
	\@citex}{\@tempswafalse\@citex[]}}
\def\@cite#1#2{{$\null^{#1}$\if@tempswa\typeout
	{IJCGA warning: optional citation argument 
	ignored: `#2'} \fi}}

\def\pmb#1{\setbox0=\hbox{#1}
	\kern-.025em\copy0\kern-\wd0
	\kern.05em\copy0\kern-\wd0
	\kern-.025em\raise.0433em\box0}

\def\fnt#1#2{\footnotetext{\kern-.3em
	{$^{\mbox{\scriptsize #1}}$}{#2}}}

\def\fpage#1{\begingroup
\voffset=.3in
\thispagestyle{empty}\begin{table}[b]\centerline{\footnotesize #1}
	\end{table}\endgroup}

\def\runninghead#1#2{\pagestyle{myheadings}
\markboth{{\protect\footnotesize\it{\quad #1}}\hfill}
{\hfill{\protect\footnotesize\it{#2\quad}}}}
\font\tenrm=cmr10
\font\tenit=cmti10 
\font\tenbf=cmbx10
\font\bfit=cmbxti10 at 10pt
\font\ninerm=cmr9
\font\nineit=cmti9
\font\ninebf=cmbx9
\font\eightrm=cmr8

\def\qed{\hbox{${\vcenter{\vbox{			
   \hrule height 0.4pt\hbox{\vrule width 0.4pt height 6pt
   \kern5pt\vrule width 0.4pt}\hrule height 0.4pt}}}$}}

\def\@refcitex[#1]#2{\if@filesw\immediate\write\@auxout  
	{\string\citation{#2}}\fi 
\def\@citea{}\@refcite{\@for\@citeb:=#2\do 
	{\@citea\def\@citea{, }\@ifundefined 
	{b@\@citeb}{{\bf ?}\@warning 
	{Citation `\@citeb' on page \thepage \space undefined}} 
	\hbox{\csname b@\@citeb\endcsname}}}{#1}} 
 \def\@refcite#1#2{{[{#1}]\if@tempswa\typeout     
        {WSPC warning: optional citation argument 
	ignored: `#2'} \fi}} 
 \def\refcite{\@ifnextchar[{\@tempswatrue 
	\@refcitex}{\@tempswafalse\@refcitex[]}}

\def\beq{\begin{equation}}
\def\eeq{\end{equation}}

\def\beqar{\begin{eqnarray}}
\def\eeqar{\end{eqnarray}}

\begin{document}
\setlength{\textheight}{7.7truein}  
\addtolength{\topmargin}{+50pt}

\runninghead{D. V. Ahluwalia}{
Interface of Gravitational and Quantum Realms
$\ldots$}

\normalsize\textlineskip
\thispagestyle{empty}
\setcounter{page}{1}

\copyrightheading{}			

\vspace*{0.88truein}

\fpage{1}
\centerline{\bf INTERFACE OF GRAVITATIONAL AND QUANTUM REALMS}
\vspace*{0.37truein}
\centerline{\footnotesize D. V. AHLUWALIA}
\baselineskip=12pt
\centerline{\footnotesize\it 
Theoretical Physics Group, Fac. de Fisica}
\baselineskip=12pt
\centerline{\footnotesize\it 
Univ. Aut. de Zacatecas, Ap-Postal C-600,
Zacatecas, ZAC 98062, Mexico}
\centerline{\footnotesize\it 
ahluwalia@phases.reduaz.mx}

\publisher{(received date)}{(revised date)}

\vspace*{0.21truein}
\abstracts{
The talk centers around the question: Can general-relativistic  
description of physical reality be considered complete? 
On the way I argue how -- unknown to many a 
physicists, even today -- 
the ``forty orders of magnitude argument'' against quantum gravity
phenomenology was defeated more than a quarter of a century ago,
and how we now stand at the possible verge of detecting a signal for the
spacetime foam, and studying the gravitationally-modified wave particle
duality using superconducting quantum interference devices. 
}{}{}

\vspace*{10pt}

\nonumsection{Preamble}
The idea for the First IUCAA Meeting on the Interface of
Gravitational and Quantum Realms (17-21 December 2001, Pune, India)
arose during a walk, a year before, with Naresh 
Dadhich. A reader who was not at IGQR-I is likely to find
a contradiction between what I write here and what Naresh Dadhich and 
I  write in the opening lines of the Preface. 
For that reader I note that the physics walks at IUCAA often begin
with a left turn exit from the main IUCAA entrance, they wind through
a narrow road on the outskirts of  an open field. 
Mid way in the walk, on the left of that narrow lane, is a tree. The tree
provides shade for a {\em chai\/} (``tea'') and 
{\em samosa} break, and hosts 
birds of several species.
After the chai the walk continues. The walk, punctuated by a 
chai and
samosa break, is an important part of life at IUCAA. Such walks
inspire a whole range of new ideas and provide an opportunity for monastic
reflections.

\def\A{\stackrel{\hspace{0.04truecm}\circ }{\mbox{A}}}


\vspace*{1pt}\textlineskip	
\section{Introduction}	
\vspace*{-0.5pt}

\noindent
The foundations of the modern gravitational and quantum frameworks were 
established in an intellectually turbulent era of the early twentieth 
century. The rapid developments of the theory of general relativity 
on the one hand, and the similarly fast evolution of the theory of  
quantum mechanics on the other -- under schools which were essentially 
opposed in their philosophical outlooks \refcite{epr,bohr} -- have, 
in my opinion, left the interface of the two frameworks largely unexplored.

The  Bell paper \refcite{bell} finally placed the concerns on the
incompleteness of quantum framework \refcite{epr} to experimental
front \refcite{ex1,ex2,ex3,ex4,ex5}. 
Yet, if it was asked for the quantum realm,  
``Can quantum-mechanical description of physical reality be considered
complete?,'' then asking, 
``Can general-relativistic  description of physical reality be considered
complete?,'' may, I suspect, lead to interesting insights into the
interface of the two rival views on physical reality.

Towards this end, in the next section I briefly review
the canonical dimensional arguments which place the interface 
of the gravitational and quantum realms (IGQR) beyond the reach
of any quantum gravity phenomenology (QGP) in  an environment
independent manner. However, since many aspects of the IGQR depend
on the gravitational environment/context, 
the canonical wisdom is no longer applicable, and I argue for a viable 
QGP program for exploring IGQR. The remainder of the talk is devoted 
to theoretical remarks on the IGQR, and suggestions 
for its exploration in terrestrial laboratories.

\section{Viability of QGP program for exploring IGQR:  Defeating the
40 orders of magnitude argument.}

\noindent
To prepare for examining the experimental viability of
exploring the IQGR, I first enumerate, for
ready reference, {\em some\/}\footnote{That is,
ignoring the relevant cosmological and astrophysical numbers.} ~of the   
relevant numbers in Table \ref{tab1}.
Using the  listed constants, in an environment-independent context, 
one readily obtains various scales at which quantum-gravity shall 
become manifest
\beqar
&&\lambda_P=\sqrt\frac{\hbar G}{c^3} = 1.62\times 10^{-33}\,\,\mbox{cm},\quad
f_P= \frac{c}{\lambda_P}=1.85\times 10^{43}\,\,\mbox{Hz},\\
&& m_P = \sqrt\frac{\hbar c}{G} = 2.18 \times 10^{-5}\,\,\mbox{g},\quad
\gamma=\frac{G m_n^2/\hbar c}{e^2/\hbar c} = 8.1 \times 10^{-37}. 
\label{forty}
\eeqar
In the canonical arguments, the smallness of Planck length, $\lambda_P$,
in comparison to atomic,\footnote{Typical atomic dimension
is of the order of $\A$ngstr\"om , $\A=10^{-8}\,\,\mbox{cm}$.
Typical 
nuclear length scale is of the order of a Fermi,  $F =10^{-13}
\,\,\mbox{cm}$.
}
and nuclear dimensions,
renders 
IGQR beyond the reach of terrestrial experiments. 
At the same time, in comparison to energies to which elementary particles
can be accelerated in high-energy physics accelerators, 
the Planck mass, $m_P$, is seen to be exceedingly large
to allow any terrestrial exploration of IGQR.
The extreme smallness of $\gamma$, similarly, encodes the same conclusions.
These arguments, owing to second of Eqs. (\ref{forty}), 
are knows as the ``40 orders
of magnitude'' disparity between realm of quantum electrodynamics and 
and the regime of quantum gravity.

\begin{table}
\begin{center}
\begin{tabular}{|l|l|l|}\hline
Proton charge & $e$ & $ 4.80\times10^{-10}\,\,\mbox{esu}$\\ \hline
Gravitational constant & $G$ & 
				  $6.67 \times 10^{-8}$ 
                                  $\,\,\mbox{dyn-cm}^2/\mbox{g}^2$ \\ \hline
Planck's constant/$2\pi$ & $\hbar$ & 

				   $1.05\times 10^{-27} \,\,\mbox{erg-s}$ \\
\hline
Speed of light in vacuum & $c$ & $3\times 10^{10}\,\, \mbox{cm/s}$ \\ \hline
Nucleon mass & $m_n$ & $1.67\times 10^{-24}\,\,\mbox{g}$\\\hline
\end{tabular}
\caption{For ready reference, 
some constants relevant for the interface of the gravitational
and quantum realms. All numbers cited here, and rest of the paper, 
are in the sense of ``$\approx$.''}
\label{tab1}
\end{center}
\end{table}

As we move through this talk, we shall see that such arguments 
are misleading. To make first of such arguments, 
in Table \ref{tab2}, I collect together some of the quantities that
define the terrestrial gravitational environment.
In conjunction with Table \ref{tab1}, it allows  
us to construct the following dimensionless
gravitational potentials:
\beq
\phi_\oplus= - \frac{G M_\oplus}{c^2 R_\oplus} = - 6.96\times 10^{-10},\quad
\phi_n=-\frac{G m_n}{c^2 F} = - 1.24\times 10^{-39}\,.
\eeq
Immediately, I take note of the circumstance that in going from
the nuclear realm to laboratories on the surface of the earth we
gain thirty orders of magnitude (towards defeating $\gamma$):
\beq
\frac{\phi_\oplus}{\phi_n}= 0.56\times 10^{30}\,,
\eeq
In addition, if we experiment with thermal neutrons with about an $\A$
wavelength, then for a $10\,\,\mbox{cm}$ table-top 
interferometer arm, we obtain
a dimensionless number, $10^9$. The interplay of 
these two large numbers comfortably
overcomes the perceived disadvantage for QGP in IGQR.

\begin{table}
\begin{center}
\begin{tabular}{|l|l|l|}\hline
Earth mass & $M_\oplus$ & $ 5.98\times 10^{27}\,\,\mbox{g}$ \\ \hline
Earth  radius [Mean] & $R_\oplus $ & $6.37\times 10^{8} \,\,\mbox{cm}$
						\\ \hline
Solar mass & $ M_\odot$ & $2 \times 10^{33} \,\,\mbox{g}$ \\ \hline
``Great Attractor region (GAR)'' mass & $M_{GAR}$ & $10^{17} M_\odot$ \\ \hline
``GAR - Milky Way (MW)''  distance & $R_{GAR-MW}$ 
& $150$ million light years,\\ 
 & & i.e. $1.42\times 10^{26}\,\,\mbox{cm}$ \\ \hline
\end{tabular}
\caption{For ready reference, 
some quantities associated with our gravitational environment.
}
\label{tab2}
\end{center}
\end{table}

In fact such arguments underlie at the heart of the 
gravitationally-induced phases {\em experimentally\/}
observed in neutron \refcite{oc,cow,ja,also} and atomic \refcite{chu} 
interferometry. I summarize  the experimental results in 
Table \ref{tab3}.
The cited experiments probe classical gravity in a
quantum context and verify equality of the inertial
and gravitational masses at different levels of accuracy.
Note that experiments in Ref. \refcite{cow,littrell} refer to neutron
interferometry, while the experiment of Ref. \refcite{chu} is in
the context of atomic interferometry. The discrepancy noted in
\refcite{littrell}, if it persists experimentally, would be a
serious challenge to the equality of the inertial and 
gravitational masses for neutron and carries serious implications, 
not only for classical theory of gravitation, but also
for quantum gravity. However, this circumstance is besides the 
point. The essential point to be made is that {\em experiment\/}
must remain the main guide and any empirical 
confirmation/surprise carries implications for classical as 
well quantum aspects of gravity -- as
the latter is not immune to the structure of the former.

\begin{table}
\begin{center}
\begin{tabular}{|l|l|l|}\hline
Year [Ref.] & Observation \\ \hline\hline
1975 \refcite{cow} & ``First verification of the principle of equivalence
 \\ 
& in the quantum limit.'' \\
 & $m_i=mg$, at $1\%$ level.\\ \hline
1997 \refcite{littrell} & $m_i\ne m_g$, a few part in $1000$ discrepancy.
\\ \hline
1999 \refcite{chu} & ``...best confirmation of equivalence principle
between \\ 
& a quantum and macroscopic object.'' \\ \hline
2000 \refcite{zouw} & Undertakes to study the discrepancy of Ref. 
\refcite{littrell}. \\ 
&  No definitive conclusion, yet.  \\ \hline
\end{tabular}
\caption{Status of equivalence principle in the {\em quantum\/} realm.}
\label{tab3}
\end{center}
\end{table}

\section{Equivalence principle in the IGQR}

When one speaks of exploring equivalence principle 
one often has in mind macroscopic
classical objects. Yet, as is clear from the brief discussion
of the previous section, since 1975 the neutron interferometry
has explored equivalence principle in the context of a single mass
eigenstate. Since 1997, there is a statistically significant signal
for a violation of equivalence principle (VEP). 
It is associated with the evolution of a neutron in 
the classical gravitational 
field of Earth. It has been discussed at some length in  Ref. 
\refcite{grg2001,plb2000}.

However, more than a decade before neutron interferometry experiments
were initiated, Schiff \refcite{lsch}, and  Morrison and Gold \refcite{mg},
considered the possible violations of equivalence principle 
in quantum contexts. Good \refcite{mlg} was perhaps the first to 
note of the interesting possibilities which {\em quantum systems
in linear superposition of different mass/energy eigenstates\/} offer
to explore IGQR. Further theoretical investigations of such interesting
quantum test particles -- with no classical counterpart -- occurred
in the context of neutrino 
oscillations \refcite{ls,grf96,grf97,grf98,prd98,gl,kk,jw,ac1,ac2}.

First dedicated experiment which studied neutrino-like 
systems\footnote{Specifically, 
an atom in linear superposition of different energy eigenstates.}
~in
the classical gravitational field of Earth was in the Stanford
laboratory of Chu and his collaborators \refcite{chu}. 
It verified the principle 
of equivalence to a few parts in $10^{9}$, and at the similar level
of accuracy established that the specific quantum test particle 
experienced same gravitationally-induced acceleration as a classical
macroscopic glass object (same acceleration to $7$ parts in $10^9$). 

In all early gedanken experiments in the gravitational realm 
the attention was invariably confined to mass eigenstates. However,
in the quantum realm -- and now with mounting evidence for neutrino 
oscillations -- one must in addition consider gedanken experiments 
which allow quantum test particles in linear    
superposition of different mass/energy
eigenstates.\footnote{Borrowing a terminology from the neutrino physics,
one may call such states to be ``flavor states.''}
~Now, if inertial
and gravitational masses are considered operationally independent,
then, for such test particles (which have no classical counterpart),
the equality of inertial and gravitational masses cannot be 
claimed beyond certain fractional accuracy. This was first noted
in Refs. \refcite{grg2001,plb2000}. These  fractional accuracies
are determined by the underlying mass eigenstates, and the mixing matrix.
As accuracy of the Stanford-like experiments improves, these 
flavor-dependent fractional accuracies will become accessible to 
experimental investigations. 

References \refcite{ls,grf96,grf97,grf98,prd98,gl,kk,jw} provide
extensive discussion of flavor oscillation clocks -- a term
I coined a few years ago to emphasize unique nature of these 
systems to investigate gravity with quantum test particles
\footnote{A few-line Erratum to Ref. \refcite{grf96}
noted ``In retrospect, this paper shows that neutrino oscillations provide
a flavor oscillation clock {\em and\/} this flavor-oscillation clock 
redshifts as required by the theory of general relativity.''}~   -- and the 
effect of different gravitational environments.
Should a violation of equivalence principle
exist, the essentially constant gravitational potential produced by 
cosmological fluctuations in  matter density would become observable
in {\em local\/} measurements of gravitationally-induced redshift of
flavor oscillation clocks. Such gravitational potentials carry a 
typical dimensionless value given by (see, Great Attractor region
in Table \ref{tab2}, cf. Ref. \refcite{ik,ga}):
\beq
\phi_{GAR} = - \frac{G M_{GAR}}{c^2 R_{GAR-MW}} = - 1\times 10^{-4}\,. 
\eeq
The significance of $\phi_{GAR}$ lies in the fact that it is 
about five orders of magnitude large than $\phi_\oplus$. Depending on
functional form of a possible VEP  -- or, quantum-induced violation
of equivalence principle, qVEP, introduced in Ref.
\refcite{plb2000} -- this could significantly amplify the local
observability of $\phi_{GA}$, and hence give us a direct observational
probe for the local distribution of cosmological matter.

\section{Gravitationally modified wave-particle duality: non-commutative
spacetime, spacetime foam, and its detectability}

In the last several years it has become increasingly 
clear that interplay
of gravitational and quantum effects destroys commutativity
of various components of the spacetime vector associated with
an event, and in addition it modifies the fundamental commutator.
In one spatial dimension, a much studied scenario\footnote{Other
scenarios may be found in Ref. \refcite{fc00}.}~ for this 
non-commutativity is captured by the modification
\refcite{fc0,fc1,fc2,fc3,fc4,fc5,fc6,fc7,fc8}:  

\def\ar{\stackrel{\hspace{0.04truecm}grav. }{\mbox{$\longrightarrow$}}}

\beq
\left[x,\,p_x\right] = 
i\hbar \quad\ar \quad\left[x,\,p_x\right] = i\hbar\left(1+ \epsilon 
\frac{\lambda_p\, p_x^2}{\hbar^2}\right),
\eeq
where $\epsilon$ is of the order of unity (set equal to unity below).
This modification of the fundamental commutator, leads to the following
representative consequences/obser\-va\-tions:

\noindent
{\bf 1.}
The de Broglie wave-particle duality is modified. 
It can be encoded in
modification of de Broglie's fundamental relation \refcite{pla2000}:
\beq
\lambda_{dB}=\frac{h}{p}
\quad \ar\quad
\lambda = \frac{\overline{\lambda}_P}
{\tan^{-1}(\overline{\lambda}_P/\lambda_{dB})},
\eeq
where $\overline{\lambda}_P$ is the Planck circumference ($=2\pi\lambda_P$).
The gravitationally-modified $\lambda$ reduces to $\lambda_{dB}$ for
the low energy regime, and saturates to $4 \lambda_P$ in the 
Planck realm. Not only does this saturation suggests that spacetime
looses operational meaning at length scales below Planck length,
but it also implies that in spacetime symmetries of quantum gravity 
there exist at least {\em two}, rather than one, invariant scales. These
are $c$, and $\lambda_P$;
or, $c$ and  $\lambda_P^\prime$ 
($\lambda_P^\prime=\epsilon^\prime \lambda_P$,
with    $\epsilon^\prime$ of the order of unity). There is already 
progress in the development for a 
relativity with two invariants ($c$  and $\lambda_P$) 
\refcite{gac_rel,mv}.

The indicated saturation of $\lambda$ to $\lambda_P$ also 
implies freezing of neutrino oscillations and disappearance
of many interference phenomena \refcite{pla2000}. These could
have important phenomenological consequences -- particularly,
for the physics of early universe.

\noindent
{\bf 2.}
The  gravitationally-modified wave-particle duality is accompanied 
with an inherent non-locality \refcite{plb1994}.
Jack Ng \refcite{conj} has conjectured that this non-locality may be related
to the holographic principle \refcite{hp1,hp2,hp3}.

\noindent
{\bf 3.}
The gravitationally-modified fundamental commutator is associated
with a space-time foam -- the QGP model of a 
non-commutative spacetime. For frequencies, $ f\ll f_P$, 
Giovanni Amelino-Camelia \refcite{gac2001,gac1999,dva1999} 
has argued -- or, at the very least has made strong plausibility 
arguments -- that
the power spectrum of strain noise
\footnote{
~~Definition: Strain, 
$h = \Delta L/L$, 
where, $\Delta L$,
is the fluctuation, say induced by gravity-waves (or, space-time
fluctuations associated with QGP's spacetime foam) 
in the relevant distance $L$.}
~~is {\em constant} and
can be approximated by
\beq
\rho_h(f) \approx \frac{\lambda_P}{c} = 5.40\times 10^{-44} 
\,\,\mbox{Hz}^{-1}.
\eeq
Interestingly, such a space-time foam induced white noise
is within the reach of currently operating, and planned, gravity
wave interferometers \refcite{gac2001,gac1999}.
In addition, the associated 
Planck-scale deformations of the dispersions relations
are good candidates for solving a host of observational anomalies
\refcite{gac_2002,ss,sc}

\noindent
{\bf 4.}
Superconducting quantum interference devices (SQUIDs), 
carry  superconducting currents  
with {\em temp\-erature-tunable\/} superconducting mass 
\beq
m_s(T)\sim \eta(T)\, N_a\, m_c\,,\label{ms}
\eeq
behaving as one quantum object (under certain circumstances).
In Eq. (\ref{ms}),  
$N_a \approx 6 \times 10^{23}\,\, \mbox{mole}^{-1}$, 
$m_c\approx 2\times 0.9 \times 10^{-27}$ gm, and $\eta(T)$
encodes fraction of the available electrons that are in
a superconducting Cooper state at temperature, $T$. 
Sufficiently below the critical
temp\-rat\-ure,   $\eta(T)$ may approach unity.
The temperature-tunable, $m_s(T)$, can easily  
compete with the Planck mass, $m_P$.
Thus, 
SQUIDs carry significant potential to probe
wave-particle duality near the Planck scale. One of 
the theoretical and experimental challenges that remains is  
to devise an experiment that invokes   $m_s(T)$ rather than  
$m_c$.\footnote{~~I thank J. L. Smith  
for arranging a discussion of this subject.}

\section{Concluding remarks:  Can general-relativistic  description 
of physical reality be considered complete?}

Having taken this walk through a variety of terrains in IGQR,
I return to the original question: 
Can general-relativistic  description of physical 
reality be considered complete? The answer is an obvious no -- however, 
the precise nature of this departure is yet to be settled. 

It is not 
the flaw of the theory of general relativity (TGR), for it was never 
formulated with quantum realm in mind. 
The founders of TGR did not devise 
gedanken experiments that used quantum test particles, or 
nor did they envisage
intrinsically quantum sources \refcite{tp1987}.
Furthermore, while non-commutative spacetime was entertained 
several decades ago \refcite{snyder}, it was realized only in recent 
years that interplay of the gravitational and quantum realms 
necessarily leads to a non-commutative -- as opposed to TGR's 
spacetime continuum  -- spacetime. 

It is abundantly clear that both the gravitational and quantum realms
suffer changes as one walks from a strictly (``Strictly,'' in the sense 
as of ``as close as practically possible,'' etc.)
gravitational to a quantum realm, and vice versa. Yet,
there are circumstances in which quantum realm enters only at the
level of a quantum test particle with no classical analog, and gravitational
realm is treated in accordance with TGR. In such circumstances,  
Can general-relativistic  description of physical 
reality be considered complete? The canonical gedanken experiments,
and the principle on which TGR is formulated, invariably invoke
local equivalence of the effects of gravity, and that of 
acceleration \refcite{class}. However, one -- in an entirely
gedanken situation -- can imagine the $M_{GAR}$ of the Great Attractor region
to be distributed in a spherically symmetric manner  
around our Milky way; and the radius of such a matter distribution 
can be visualized to be $R_{GAR}$. 
This hypothetical $(M_{GAR}, R_{GAR-MW})$ configuration does not
induce any gravitational forces
because $\phi_{(M_{GAR}, R_{GAR-MW})} = -1 \times 10^{-4}$ is 
constant (i.e. is gradient-less inside the hypothetical matter
distribution).
Yet, it is responsible for inducing
gravitational redshift of 
flavor-oscillation clocks via {\em non-zero\/} gravitationally 
induced {\em relative\/} phases between the underlying mass eigenstates 
of neutrinos. These relative phases, incidentally, are precisely those
which redshift the flavor-oscillation clocks. 
Now, if one was to think away the Milky way (keeping the
neutrinos from distant sources streaming), one is left
with a region of spacetime of vanishing Riemann curvature  -- and 
the gravitational redshift of flavor-oscillation clocks. For a single
mass eigenstate these phases are global and hence unobservable.
But, when {\em different\/} mass eigenstates are superimposed
these phases become relative and hence observable. So -- while
purists would not find it difficult to counter this argument
via redshift of planetary orbits -- we do have a situation where 
gravitationally induced quantum mechanical phases exist in a situation
where the Riemann curvature identically vanishes. 
That is, a region with  vanishing Riemann curvature cannot be 
considered devoid of gravitational effects, and gravitational fields.
While all gravitationally induced forces vanish in such a region
of spacetime, relative gravitationally-induced phases in 
flavor-oscillation clocks do not.
The canonical wisdom captured in Synge's classic on TGR 
\refcite{classic}, which reads:
\begin{quote}
The essence of Einstein's general theory lies in the assumption
that gravitation manifests itself in the curvature of Riemannian 
space-time. If the Riemann tensor $R_{ijkm}$ of the metric $\ldots$
were to vanish, we would be back in the flat space-time of 
gravitationless special relativity. In fact, we may write 
symbolically 
\beq
R_{ijkm} = \mbox{gravitational field},\nonumber
\eeq
\end{quote}
is, therefore, challenged. Gravitation resides beyond
Riemann curvature in the spacetime metric. Spacetime
metric inside our hypothetical region, and one in free
fall outside this region, are not identical.  This
fact can be ascertained observationally by any good 
astronomer.

The situation becomes
even more interesting if one considers the 
fuzzy-spin gravitino of Ref. \refcite{plb2002} and considers 
its scattering from a gravitating source \refcite{rainbow}.
In this situation, as in many other, many of the proclaimed 
statements on the {\em geometrical\/} nature of gravitation 
fall apart because the founding fathers of TGR had not 
envisioned that the theory resulting from the stated principle 
contained Lens-Thirring gravitational field. 
With this  intellectual provocation I leave the remaining ponderings 
for our audiences' entertainment without further comment.

We began, with the question: 
Can general-relativistic description of physical 
reality be considered complete?
The answer which seems to emerge is: No, but precise nature of this
``no'' is yet to be settled. It is there in those arguments, in those
ponderings, that the IGQR and QPG offer some of the greatest fun.

May the phases be with you!

\nonumsection{Acknowledgments}

My thanks to Naresh Dadhich, and to Jayant Narlikar,
for making the meeting a wonderful and truly joyful
event. The laughter, the provocative presentations,
questions -- serious, and those most important ones in
good humor -- all stand as a reality dynamically 
frozen in time, and space of one week. Parampreet Singh
attended to many things with humor, and  friendship,
and to him my thanks, as to all others who made the meeting
such a good and useful experience. Special thanks, on behalf
of all Indians, to those who came from places far away 
and beyond boundaries of India. 
This work is supported by CONACyT (Mexico) Project 32067-E -- 
to CONACyT my thanks.

\nonumsection{References}
It is impossible to do justice to all the works that touch on the 
subject of this talk. The general references that I have 
collected here should help the reader delve deeper into the 
subject. After this talk was given another 
experiment in IGQR was reported. It explores quantum states 
of a neutron in Earth's gravity \cite{another}. 
Several seminal works somehow did not weave with
the flow of thoughts presented in this talk, 
and yet they are so important to IGQR that
we record one of them here as a gesture of thanks 
to its author for starting it all \refcite{sh}.

\end{document}